\documentclass[%
reprint,
showpacs,
amsmath,amssymb,
aps,prl,
]{revtex4-1}
\usepackage{booktabs}
\usepackage{tabularx}
\usepackage{braket}
\usepackage{float}
\usepackage{graphicx}% Include figure files
\usepackage{dcolumn}% Align table columns on decimal point
\usepackage{bm}% bold math
\usepackage{subfigure}
\usepackage{color}

\usepackage[colorlinks,citecolor = blue, linkcolor=red,hyperindex,CJKbookmarks]{hyperref}

\begin{document}
%\preprint{}

\title {Multiple states in turbulent large-aspect ratio thermal convection: \\
What determines the number of convection rolls?} 

\author{Qi Wang$^{1,2}$}
\author{Roberto Verzicco$^{4,5,1}$}
\author{Detlef Lohse$^{1,3}$}\email{d.lohse@utwente.nl}
\author{Olga Shishkina$^{3}$}\email{Olga.Shishkina@ds.mpg.de}
\affiliation{$^1$Physics of Fluids Group and Max Planck Center for Complex Fluid Dynamics, MESA+ Institute and J. M. Burgers Centre for Fluid Dynamics, University of Twente, P.O. Box 217, 7500AE Enschede, The Netherlands\\
$^2$Department of Modern Mechanics, University of Science and Technology of China, Hefei 230027, China\\
$^3$Max Planck Institute for Dynamics and Self-Organization, 37077 G\"ottingen, Germany\\
$^4$Dipartimento di Ingegneria Industriale, University of Rome `Tor Vergata',
Via del Politecnico 1, Roma 00133, Italy\\
$^5$Gran Sasso Science Institute - Viale F. Crispi, 767100 L'Aquila, Italy.}

%\thanks{}

\date{\today}

%\pacs{47.20.Ky, 46.40.Jj, 47.63.Gd, 87.85.gf}
%\keywords{}

\begin{abstract}

% Now 594 symbols! Must be \le 600
%For the same control parameters, wall-bounded flows can take different statistically stationary turbulent states.
%Thus, in 2D turbulent Rayleigh--B\'enard flows, the heat and momentum transport depends on the aspect ratio $\Gamma_r$ of the realized convection rolls, being more efficient for slender rolls (small $\Gamma_r$). Viscous damping and elliptical instability determine a $\Gamma_r$-window for the realizable states; what state the system takes depends on the initial conditions. 
%Theoretically derived $\Gamma_r$-frames for no-slip  and free-slip boundary conditions are in excellent agreement with our numerical data. 
%

%% for counter: 
%For the same control parameters, wall-bounded flows can take different statistically stationary turbulent states. Thus, in 2D turbulent Rayleigh-Benard flows, the heat and momentum transport depends on the aspect ratio Gr of the realized convection rolls, being more efficient for slender rolls (small Gr). Viscous damping and elliptical instability determine a Gr-window for the realizable states; what state the system takes depends on the initial conditions. Theoretically derived Gr-frames for no-slip  and free-slip boundary conditions are in excellent agreement with our numerical data. 

%1300 symbols or so. Must be \le 600
Recent findings suggest that wall-bounded turbulent flow can take different statistically stationary turbulent states, with different transport properties, even for the very same values of the control parameters. What state the system takes depends on the initial conditions. Here we analyze the multiple states in large-aspect ratio ($\Gamma$) two-dimensional turbulent Rayleigh--B\'enard flow with no-slip plates and horizontally periodic boundary conditions as
model
system. % In particular, 
We determine the number $n$ of convection rolls, their mean aspect ratios $\Gamma_r = \Gamma /n$, and the corresponding transport properties of the flow (i.e., the Nusselt number $Nu$), as function of the control parameters Rayleigh ($Ra$) and Prandtl number. The effective scaling exponent $\beta$ in $Nu \sim Ra^\beta$ is found to depend on the realized state and thus $\Gamma_r$, with a larger value for the smaller $\Gamma_r$. 
By making use of a generalized Friedrichs inequality, we 
show that  the elliptical instability and viscous damping determine 
the  $\Gamma_r$-window for the realizable turbulent states. 
%and 
%theoretically show that statistically stationary turbulent states can only have convection rolls with aspect ratios in a certain window, 
%%%$0.73 \le \Gamma_r \le 1.36$, 
The theoretical results are 
in  excellent agreement with our numerical finding $2/3 \le \Gamma_r \le 4/3$, where the lower threshold is approached for the larger $Ra$. Finally, we show that the theoretical approach to  frame $\Gamma_r$ also works for free-slip boundary conditions. 
\end{abstract}

\maketitle

%\section{}
%\subsection{}
%\subsubsection{}

For laminar flows, flow transitions can often be calculated from linear stability analysis. 
Such an analysis not only gives the critical value of the control parameter at which the instability sets in, but also the wavelength of the emerging structure.
Famous classical examples for linearly unstable wall-bounded flows are the onset of convection rolls in Rayleigh-B\'enard convection or the onset of Taylor rolls in Taylor-Couette flow \cite{dra81}. 
In both cases, the rolls of the most unstable mode have a certain wavelength which follows from the linear stability analysis. With increasing flow driving strength, more and more modes become unstable, and in the fully turbulent case the base flow is unstable to basically any perturbation. 

What then sets the size of the flow structures in such {\it fully turbulent} wall-bounded flow? 
Recent findings have suggested that wall-bounded turbulent flows can take {\it different} statistically stationary turbulent states, with different length scale of the flow structures and with different transport properties, even for the very same values of the control parameters. Examples for the coexistence of such multiple turbulent states  include turbulent (rotating) Rayleigh-B\'enard convection \cite{xi08, poe11, poe12, weiss2013, wang2018, xie2018, favier2019}, Taylor-Couette turbulence \cite{hui14, veen2016, ost16}, von Karman flow \cite{rav04, rav08, cor10, faranda2017}, rotating spherical Couette flow \cite{zim11}, Couette flow with span-wise rotation \cite{xia2018}, but also geophysical flows \cite{bouchet2009random, bouchet2012statistical} such as in ocean circulation \cite{broecker1985, schmeits2001, ganopolski2002}, in the liquid metal core of Earth \cite{glatzmaiers1995, li2002, olson2010, sheyko2016magnetic}, or in the atmosphere
\cite{weeks1997, bouchet2019rare}. 

The occurrence of multiple states in fully turbulent flows  can be considered unexpected since, according to Kolmogorov \cite{kol41b}, for strong enough turbulence, the fluctuations should become so strong that the whole highly dimensional phase-space is explored.
Of course, one could argue that in the above given cases and examples, the turbulence driving has not yet been strong enough to reach that state and  that the occurrence of multiple turbulent states in wall-bounded turbulence may be a finite size effect, but in any case even then it remains open what sets the range of allowed sizes of the flow structures in such turbulent flows.

\begin{figure}
\includegraphics[width=0.45\textwidth]{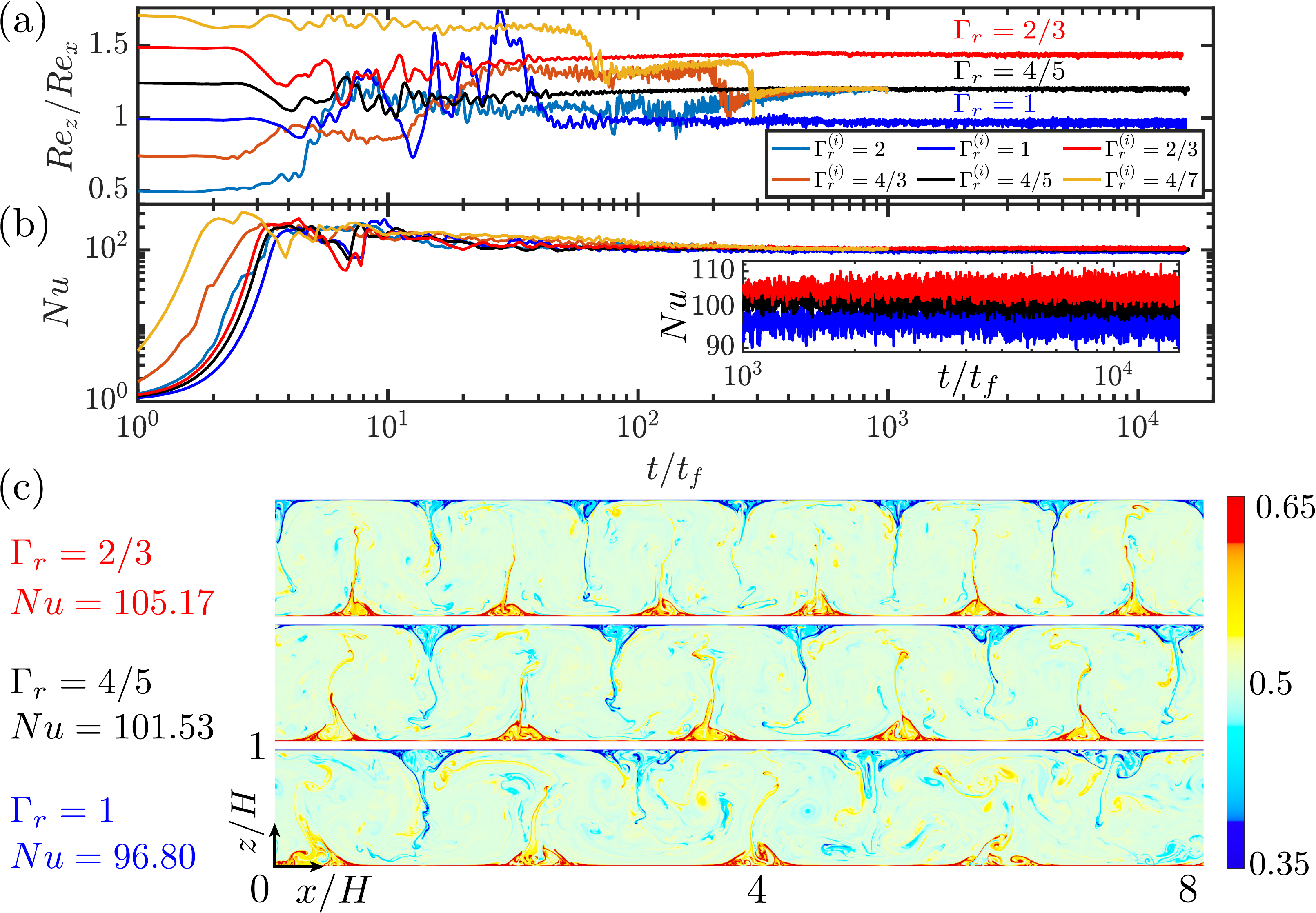}
\caption{
(a, b) Time evolution of (a) $Re_z/Re_x$ and (b) $Nu$ for different initial roll states, $Ra=10^{10}$, $Pr=10$, $\Gamma=8$.
$\Gamma_r^{(i)}=\Gamma / n^{(i)}$ is the initial and  $\Gamma_r=\Gamma / n$ the final mean aspect ratio of the rolls; $n^{(i)}$ is the initial and  $n$ the final number of the rolls.
Note the logarithmic scales of the time axes.
%(c) $Re_z/Re_x$ as a function of $\Gamma_r$, for all studied combinations of $Ra$, $Pr$ and $\Gamma=8$. 
%The dashed curve follows $\Gamma_r^{-1}$.
(c) Snapshots of the temperature fields for the three statistically stable turbulent states 
for $\Gamma_r=1$ ($n=8$), $\Gamma_r=4/5$ ($n=10$) and  $\Gamma_r=2/3 $ ($n=12$) and the corresponding Nusselt numbers, for $Ra=10^{10}$, $Pr=10$, and $\Gamma=8$.
All sub-figures are for no-slip BCs.
}
\label{fig1}
\end{figure}

To illuminate this question, as a model system we picked two-dimensional (2D) Rayleigh--B\'enard (RB) turbulence, the flow in a closed container heated from below and cooled from above \cite{ahl09, loh10, chi12}. 
The control parameters are the Rayleigh number $Ra$, which is the dimensionless temperature difference between the plates, the Prandtl number $Pr$, which is the ratio between kinematic viscosity ($\nu$) and thermal diffusivity ($\kappa$), and the aspect ratio $\Gamma=W/H$, i.e., the ratio between horizontal ($W$) and vertical ($H$) extension of the system. 
The response parameters are the Nusselt number $Nu={QH}/{(k\Delta})$ and the Reynolds number $Re={UH}/{\nu}$, which indicate the dimensionless heat transport and flow strength in the system. 
Here $Q$ is the heat flux crossing the system, $k$ the thermal conductivity, $\Delta$ the temperature difference at the plates, and 
%$U=\sqrt{\left<\boldsymbol{u}\cdot\boldsymbol{u}\right>_{V,t}}$ 
{$U=\sqrt{\langle{\bf u}^2\rangle_{V,t}}$}
the time and volume-averaged velocity.

The flow dynamics is given by the Boussinesq approximation of the Navier-Stokes and 
the advection-diffusion equation, with the corresponding boundary conditions (BCs) for the 
temperature and velocity  fields. 
For the latter at the plates we will first apply no-slip 
BCs, but later also examine free-slip  
BCs -- a difference which will turn out to be major for the range of allowed states.  
Periodic 
BCs are used in the horizontal direction.

We are very much aware of the differences between 2D and 3D RB flow \cite{poe13}, but in particular for large $Pr \ge 1$ there are extremely close similarities between 2D and 3D RB flows, and we wanted to pick a model system for which (i) we can reasonably explore the considerable parameter space for a large enough number of initial flow conditions and (ii) we have the chance to obtain exact analytical results for the range of allowed flow structures.

 The parameter range we will explore is for large Prandtl numbers in the range $1\le Pr \le 100$, for Rayleigh numbers in the range $10^7 \le Ra \le 10^{10}$ and for large $\Gamma$ up to $\Gamma = 32$. Note that in 2D RB, multiple coexisting  turbulent states had been found before for $Ra = 10^7$, $Pr=0.7$ and $\Gamma \approx 0.64$ (i.e., in an extremely limited range of the parameter space)~\cite{poe11}, but not for such large $\Gamma$  and $Ra$ as we explore here, as the range of chosen 
initial flow conditions was not large enough \cite{poe12}, and clearly not as general and omnipresent as we will find here.  

The direct numerical simulations were done with an advanced finite difference code (AFiD~\cite{poe15cf}) with the criteria for 
the grid resolution, as found to be required in ref.\ \cite{shi10}. The code has extensively been tested and benchmarked against other codes \cite{kooij2018} and applied in 2D RB even up to very large $Ra = 4.64\times10^{14}$ \cite{zhu18b,zhu2019reply}. More simulation details for all explored cases can be found in the supplementary material.
In order to trigger the possible convection roll state, we use different initial roll states generated by a Fourier basis: $u(x,z)={\rm sin}(n^{(i)}\pi x/\Gamma){\rm cos}(\pi z)$,
$w(x,z)=-{\rm cos}(n^{(i)}\pi x/\Gamma){\rm sin}(\pi z)$,
 where $n^{(i)}$ indicates the initial number of rolls in the horizontal direction. The initial temperature has a linear profile with random perturbations.

In Fig.~\ref{fig1}(a) and ~\ref{fig1}(b) we show the temporal evolution of some flow characteristics for six  different initial flow conditions for the case of $Ra = 10^{10}$, $Pr= 10$, and  $\Gamma = 8$. 
We vary the initial number $n^{(i)}$ of rolls from $n^{(i)}=4$ to $n^{(i)} = 14$, implying aspect ratios of the initial rolls from $\Gamma_r^{(i)} = \Gamma /n^{(i)} = 2$ to 
$\Gamma_r^{(i)}=4/7$.
As flow characteristics we picked the Reynolds number ratio $Re_z/Re_x$ and the Nusselt number $Nu$. Here $Re_z = \sqrt{\langle w^2\rangle_V} H/\nu$ is  the volume averaged vertical Reynolds number and $Re_x = \sqrt{\langle u^2\rangle_V} H/\nu$ the horizontal one, where
$w(t)$ and $u(t)$ are the respective velocities.
As one can see in Fig.~\ref{fig1}(a, b), depending on the six initial conditions, over the very long time of more than $10^4$ free-fall time units, the system evolves to either of three different final turbulent states, all with different Reynolds number ratio $Re_z/Re_x$ and Nusselt number $Nu$. 
The smaller the final mean aspect ratio $\Gamma_r$ of the rolls, the larger the Reynolds number ratio $Re_z/Re_x$ and $Nu$, due to more plume-ejecting regions which have strong vertical motion. 
% Based on the data for $\Gamma=8$, -- is info in caption. not needed here. not essential. 
%In Fig.~\ref{fig1}(c) we show that $Re_z/Re_x \sim \Gamma_r^{-1}$, independent on the specific values of 
%$(Ra,Pr)$. We will explain this dependence later in the manuscript. 
%%%, which simply follows from incompressibility({\color{red}{Is it straightforward to arrive this conclusion? For free-slip cases we cautually observe $Re_z/Re_x\sim \Gamma_r^{-0.68}$}}).
%Mass conservation and incompressibility imply $Re_x=\Gamma_r Re_z$ and this is supported by the numerical data, as one can see in Fig.~\ref{fig1}(c).

The time evolution of some of the six different initial states ($n^{(i)}=6,14$) analyzed in Fig.~\ref{fig1}(a, b) can be seen in the supplementary movies, displaying  roll merging and splitting events. 
The states with large initial rolls (corresponding to $n^{(i)}=4$,~6) break up quickly, while those with  small initial rolls ($n^{(i)}=14$) first 
undergo a transition 
into an unstable twelve-roll state (with smaller $Re_z/Re_x$ than the stable one) as seen in Fig.~\ref{fig1}(a), 
followed by  merging events of two neighboring convection rolls.
Though both, the Reynolds number ratio and the Nusselt number, keep on fluctuating in time, reflecting the turbulent nature of the states, the three final statistically stable turbulent states are clearly distinguished. We characterize them by the final aspect ratio of their rolls, namely $\Gamma_r = 1$, 
$\Gamma_r = 4/5$, 
and $\Gamma_r = 2/3$, corresponding to $n= \Gamma/\Gamma_r = 8$, 10, and 12 rolls, respectively. Snapshots of these states and their corresponding Nusselt numbers are shown in Fig.~\ref{fig1}(c). 
As one can see, the larger the number of rolls, the better the (heat)  transport properties of the system, a characteristics which was found in Taylor-Couette flow before \cite{hui14} and which can intuitively be understood, due to the larger number of emitted plumes at the interfaces between the rolls.

\begin{figure}
\includegraphics[width=0.45\textwidth]{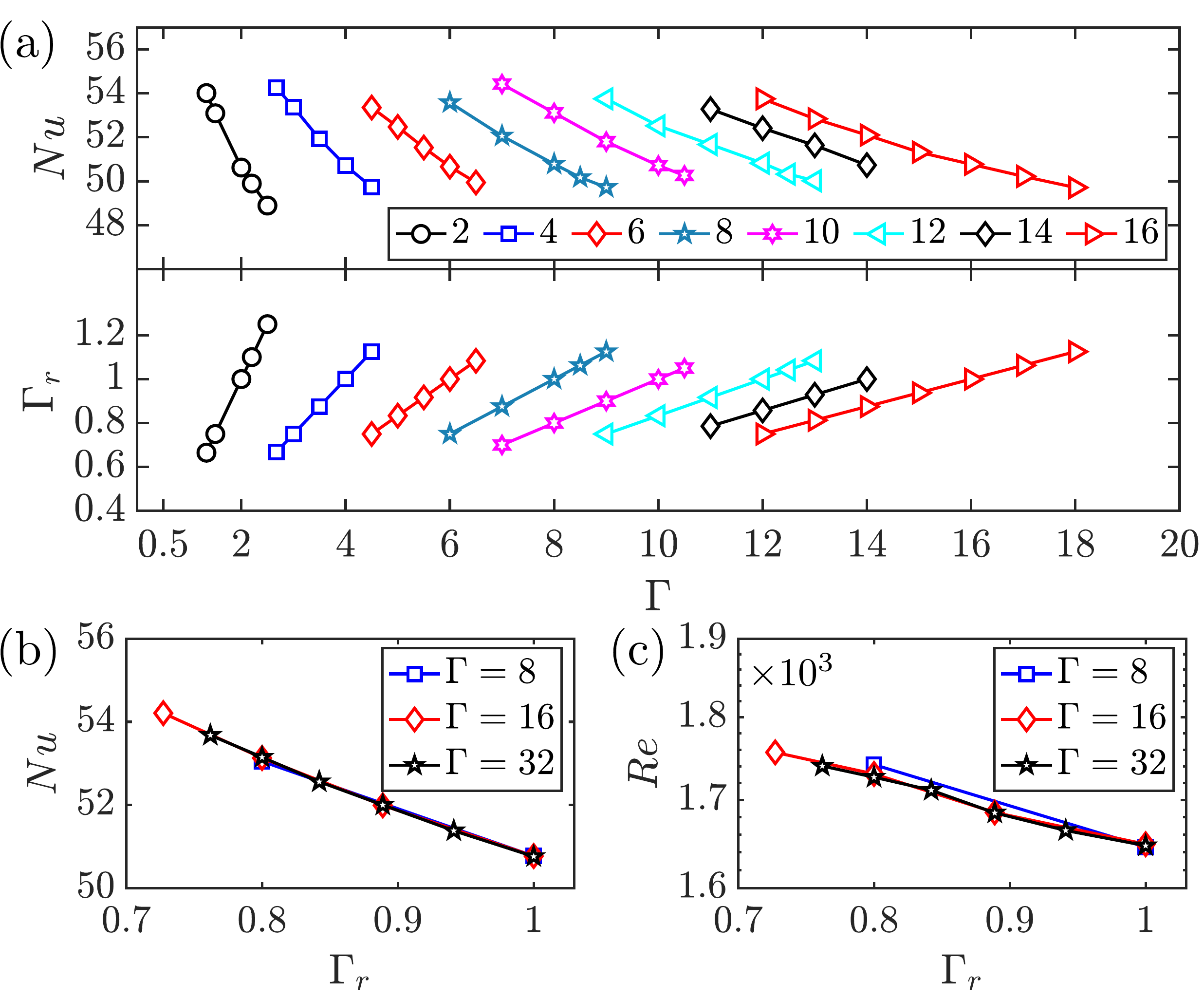}
\caption{(a) $Nu$ and (b) final aspect ratio  $\Gamma_r = \Gamma / n$ of individual rolls, as function of $\Gamma$, for different turbulent states. The numbers and colors
 in the legend denote the number $n$ of convection rolls of that state.
(c) $Nu$ and (d) $Re$, as  functions of $\Gamma_r = \Gamma /n$ for three different values of $\Gamma$. 
In this figure  $Ra=10^{9}$ and $Pr=10$ and the BCs are no-slip. }
\label{fig2}
\end{figure}

Consequently, when the cell aspect ratio $\Gamma$ is stretched, the stretching of the mean aspect ratio $\Gamma_r = \Gamma /n$  of the corresponding individual rolls is accompanied with a decrease of the corresponding Nusselt number, as seen in Fig.~\ref{fig2}(a) and \ref{fig2}(b). 
Though this behavior has been seen before \cite{poe12}, in Fig.~\ref{fig2}(a) and \ref{fig2}(b) we clearly see the coexistence of the different turbulent states. 
The determining relevance of the final mean aspect ratio $\Gamma_r = \Gamma /n $ of the individual rolls for the Nusselt number $Nu$ and Reynolds number $Re$  in the statistically stationary case is nicely demonstrated in Fig.~\ref{fig2}(c) and \ref{fig2}(d), where we show that the dependences $Nu(\Gamma_r)$ and $Re(\Gamma_r)$ are universal and irrespective of the individual values of $\Gamma $ or $n$.

\begin{figure}
\includegraphics[width=0.45\textwidth]{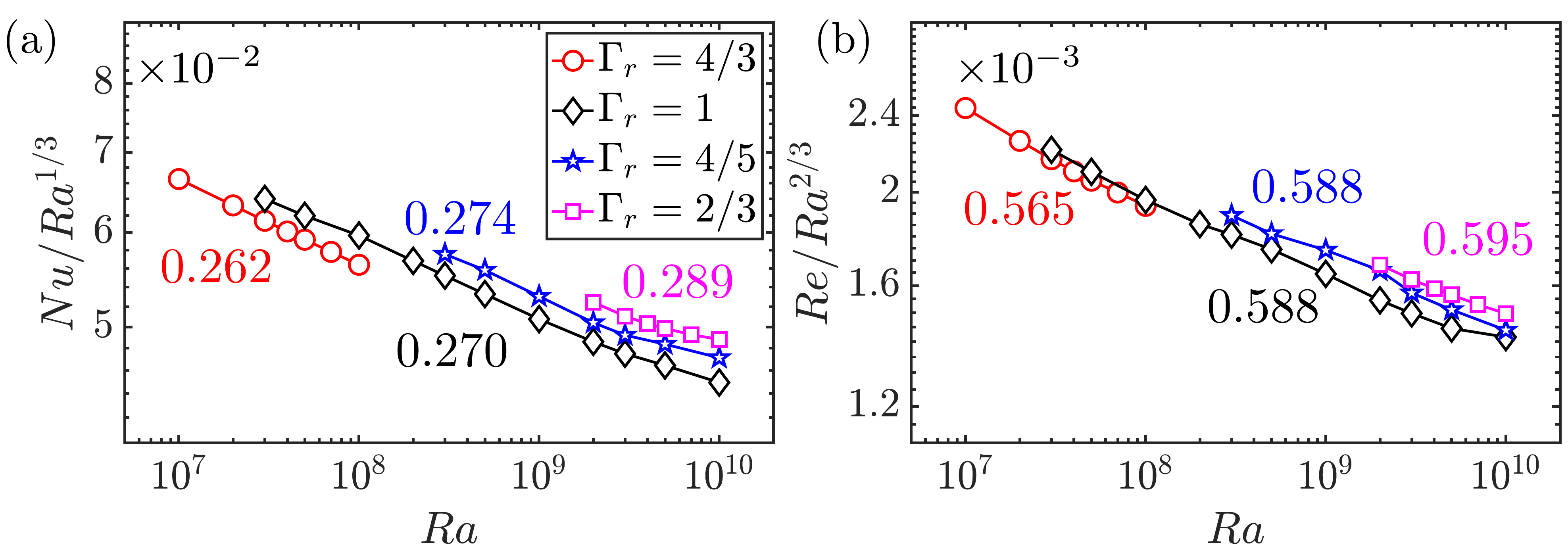}
\caption{(a) Compensated Nusselt number $Nu/Ra^{1/3}$ and (b) compensated Reynolds number $Re/Ra^{2/3}$, as functions of $Ra$, for four different turbulent states as characterized by $\Gamma_r$,
for $Pr=10$, $\Gamma=8$, and no-slip BCs. 
The effective scaling exponents $\beta$ and  $\gamma$ 
 in the scaling relations $Nu \sim Ra^{\beta}$ and  $Re \sim Ra^{\gamma}$ are shown next to the curves
 in the respective color 
 of the state and curve. 
 % for different roll states as characterized by $\Gamma_r$, 
%(c) Nusselt number $Nu$ and (d) compensated Reynolds number $Re/Pr^{-1}$, as function of $Pr$, for two different turbulent states for $Ra=10^{9}$, $\Gamma=8$, and no-slip BCs.
}
\label{fig3}
\end{figure}

Remarkably, not only the absolute value of the Nusselt number depends on $\Gamma_r$, but even the effective {\it scaling} behavior of $Nu$ with $Ra$, as can be seen in Fig.~\ref{fig3}(a). %  and Table \ref{tab01}. 
The same holds for the Reynolds number, Fig.~\ref{fig3}(b).  
In both cases the effective scaling exponent is larger for turbulent states with smaller mean
aspect ratio $\Gamma_r$ of the rolls (see the values given in Fig~\ref{fig3}(a,b)),
 i.e., when the system can accommodate a larger number $n = \Gamma / \Gamma_r$ of rolls, presumably reflecting the larger disorder and the larger number of emitted plumes for those states.

\begin{figure}
\includegraphics[width=0.45\textwidth]{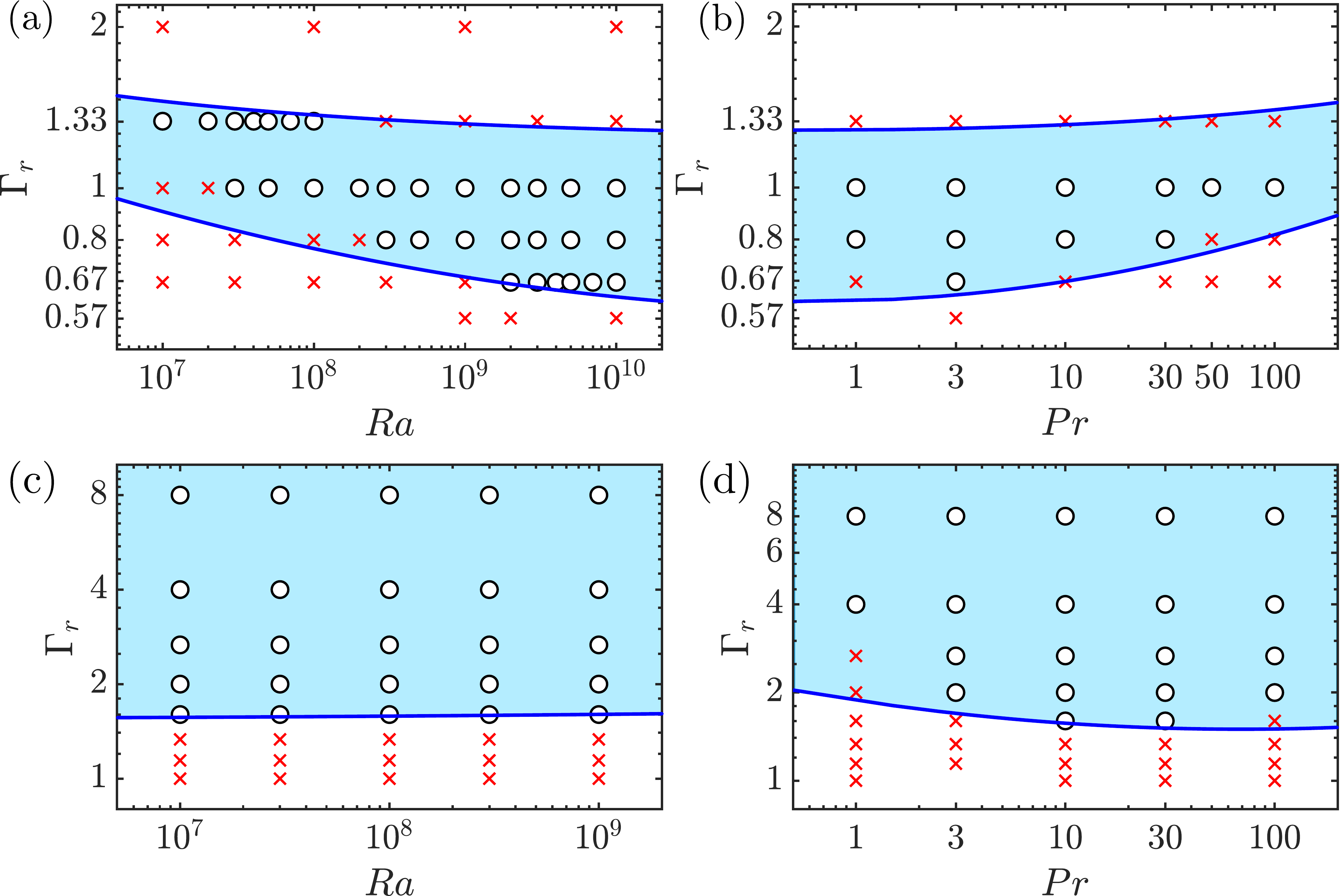}
\caption{Phase diagram in the (a, c) $Ra-\Gamma_r$ and (b, d) $Pr-\Gamma_r$ parameter space
{for (a, b) no-slip BCs and (c, d) free-slip BCs:}
%%The top row for no-slip BCs and the bottom row for free-slip BCs. 
(a) $Pr=10$, $\Gamma=8$; (b) $Ra=10^9$, $\Gamma=8$; 
(c) $Pr=10$, $\Gamma=16$; (d) $Ra=10^8$, $\Gamma=16$. 
Black circles denote that the corresponding roll state is stable, whereas red 
crosses mean that it  is not stable. 
The theoretical estimates for the transitions between the regimes are shown as solid lines. 
}
\label{fig5}
\end{figure}

From Fig.~\ref{fig3}(a,b) we also see that turbulent states with a too large aspect ratio $\Gamma_r$ of their rolls cease to exist with increasing $Ra$. 
Which turbulent states -- as characterized by the mean aspect ratio $\Gamma_r$ of their rolls --
are statistically stable  for given $Ra$ and $Pr$ can be seen from the phase diagrams in Fig.~\ref{fig5}.
For  fixed $Pr = 10$, all statistically stable turbulent states
in the no-slip case
have an aspect ratio $\Gamma_r$ in the range $2/3\le\Gamma_r\le4/3$, in the $Ra$-range analyzed in this paper. 
With increasing $Ra$, we see the range moving towards smaller values of $\Gamma_r$; e.g. 
with $1 \le \Gamma_r \le 4/3$ for $Ra=10^8$ and 
%%$4/5 \le \Gamma_r \le 1$ 
{$2/3\le\Gamma_r\le1$}
for $Ra= 10^{10}$, see figure \ref{fig5}(a). 
For $Ra= 10^9$, we find $2/3 \le \Gamma_r \le 1$ for all $Pr$ analysed in this paper, see Fig.~\ref{fig5}(b).

\begin{figure}
\includegraphics[width=0.45\textwidth]{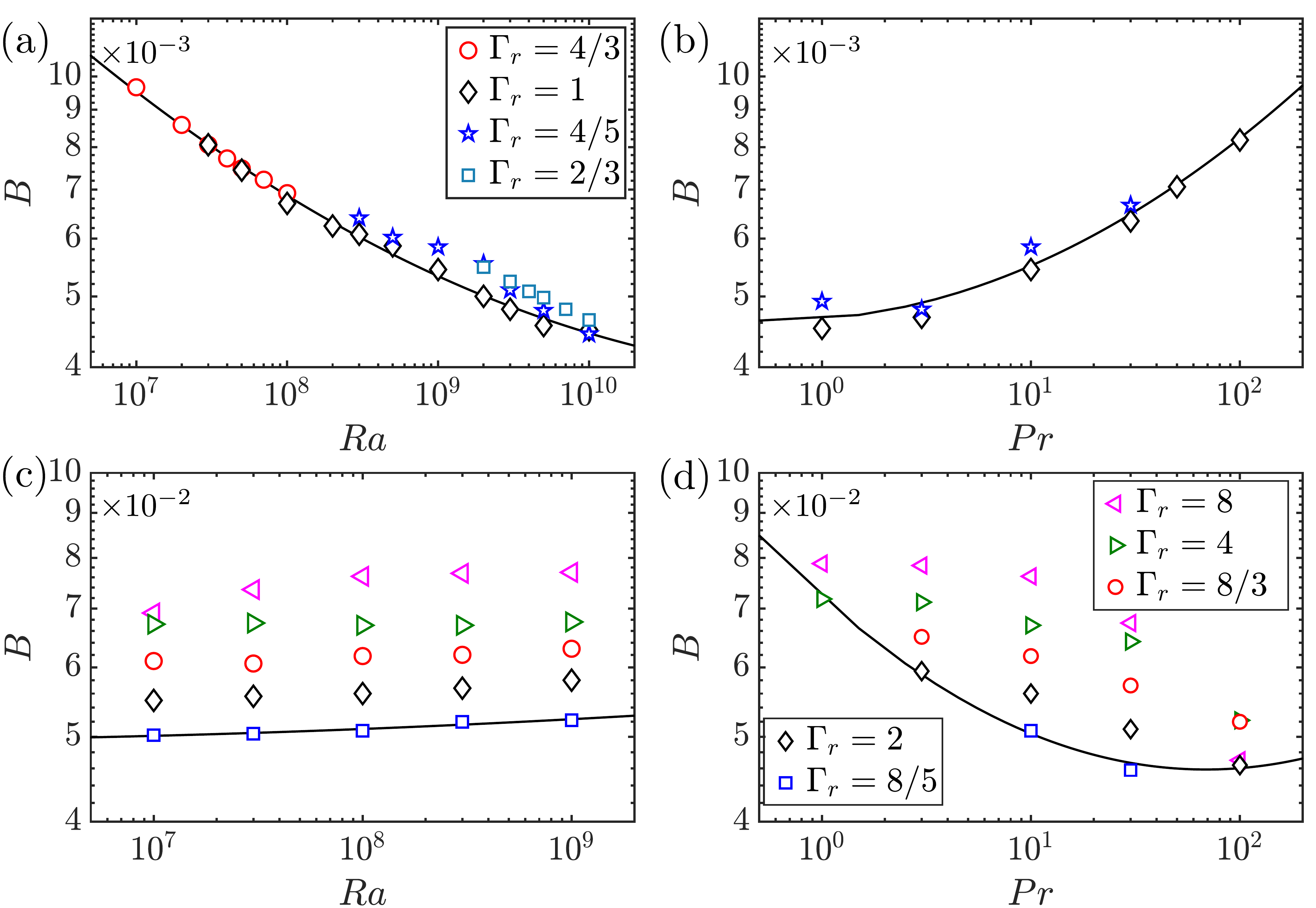}
\caption{$B={Re^2\,Pr ^2\over Ra(Nu-1)}$, as functions of (a, c) $Ra$ and (b, d) $Pr$,
 for (a, b) no-slip BCs and (c, d) free-slip BCs.
%The top row for no-slip BCs and the bottom row for free-slip BCs. 
The solid lines are fits to the data (see supplementary material). 
%with different convection roll aspect ratio $\Gamma_r$. 
%%(a) $Pr=10$, $\Gamma=8$, 
%fit for $\Gamma_r=4/3$ and 1 with 
%%$B=0.75Ra^{-0.38+1.61\times10^{-2}{\rm{log_{10}}}(Ra)}$; 
%%(b) $Ra=10^9$, $\Gamma=8$, 
%fit for $\Gamma_r=4/5$ and 1 with 
%%$B=4.65\times10^{-3}Pr^{2.22\times10^{-2}+5.08\times10^{-2}{\rm{log_{10}}}(Pr)}$; 
%%(c) $Pr=10$, $\Gamma=16$,
%fit for $\Gamma_r=1.6$ with 
%%$B=5.46\times10^{-2}Ra^{-1.69\times10^{-2}+1.65\times10^{-3}{\rm{log_{10}}}(Ra)}$; 
%%(d) $Ra=10^8$, $\Gamma=16$, 
%fit for the smallest $B$ at each $Pr$ with 
%%$B=7.22\times10^{-2}Pr^{-0.22+5.84\times10^{-2}{\rm{log_{10}}}(Pr)}$. 
%%In (c, d), for any considered $Ra$ and $Pr$ only the smallest values of $B$ are picked for the fits.
Note that the values of $B$ in the no-slip case (a, b) are much smaller than those in the free-slip case (c, d).
}
\label{fig4}
\end{figure}

We now set out to mathematically understand the range of $\Gamma_r$ the system can take for given control parameters. 
First, we recall that the roll 
%formation is determined 
size is restricted
by the elliptical instability of the flow \cite{landman1987,waleffe1990,kerswell2002,zwirner2020}. 
%\red{(We may have to elaborate why this also holds in 2D -- )}
%We therefore assume that the essence of the flow  is described with elliptical rolls which have the stream function
We therefore assume that the essence of the flow is a set of elliptical rolls, each of which can be described by a stream function
$\Psi(x,z) = (\xi + \eta ) {z^2 \over 2} + (\xi - \eta ) {x^2 \over 2}$ with $\xi \ge \eta$. 
The aspect ratio of the rolls, $\Gamma_r$,
 is directly related to the strain $\eta$ and (half of) the vorticity $\xi$ through the relation 
$\Gamma_r = \sqrt{(\xi + \eta)/(\xi - \eta)}$,
 corresponding to 
$ \eta/\xi = (\Gamma_r^2 -1 )/(\Gamma_r^2+1)$. 
Averaging $u^2$ and $w^2$ over the area 
$[-\Gamma_rH/2,\,\Gamma_rH/2]\times[-H/2,\,H/2]$, where 
$u(x,z)=\partial \Psi/\partial z$ and $w(x,z)=-\partial\Psi/\partial x$, we obtain $Re_x^2=\Gamma_r^2Re_z^2$, which is in agreement with the simulations, see supplementary material.

As shown in refs.~\cite{landman1987,waleffe1990}, 
the growth rate $\sigma=\sigma(\eta/\xi)$ of the elliptical instability is $\sigma \approx 9\eta /16$ 
for small $\eta/\xi$ 
%% \red{(why can we assume this here?) --- as \xi \ge \eta}, 
and achieves its maximum $\sigma_{\max}\approx0.35\xi$ at $\eta/\xi\approx0.8$.
Fourier modes of the kinetic energy $E$ of the perturbations follow
$\sigma E=dE/dt$ ($=\partial E/\partial t+{\bf u}\cdot\nabla E$)
and their averages ($\overline{\cdot}$)
in time and over the roll core 
$\sigma \overline{E}\lesssim(\sqrt{\overline{{\bf u}^2}}/H)\overline{E}$, implying that
the growth rate of the instability is bounded by the mean velocity of the carrying flow, i.e.,
\begin{equation}
\sigma\lesssim (\nu/H^2)Re.
\label{sigma} 
\end{equation}

We now  make use of the exact global balance for the total enstrophy $\Omega^2$ and the mean 
kinetic
energy dissipation rate $\epsilon_u$~\cite{shr90}, namely, $4 \nu  \xi^2 = \nu \Omega^2 = \epsilon_u = \nu^3 H^{-4} (Nu-1) Ra Pr^{-2}$. 
With the  definition
\begin{equation}
 \label{B}
B\equiv {Re^2\,Pr^2 Ra^{-1}(Nu-1)^{-1}}
 \end{equation}
 (which can be seen as non-dimensionalized ratio 
between the kinetic energy and the energy dissipation rate) 
this leads to
$\sigma_{\max}\approx0.35\xi
=0.175\nu H^{-2}Re/\sqrt{B}$, implying that 
 relation (\ref{sigma}) is fulfilled when $\sqrt{B}>0.175$ (i.e., when $B\gtrsim0.03$).
For smaller $B$, from Eq.~(\ref{sigma}) and the relation between the ratio $\eta/\xi$ and $\Gamma_r$,  we obtain  
$Re\ge\frac{9}{16}\eta \frac{H^2}{\nu}=
\frac{9}{32}(\Gamma_r^2 - 1)/(\Gamma_r^2 + 1)
\sqrt{Ra(Nu-1)}/ Pr$,
which gives an estimate of the maximal size of the rolls as 
 \begin{equation}
\Gamma_r \le
\sqrt
{9+32\sqrt{B}} 
 / 
\sqrt{9-32\sqrt{B}}.
 \label{1}
 \end{equation}
%with $B\equiv {Re^2\,Pr^2 \over Ra(Nu-1)}$.
Figs.~\ref{fig4}(a, b) show the $Ra$- and $Pr$-dependences of $B$, obtained numerically for the no-slip BCs, as well as their data fits (see 
%supplement 
supplementary
material).
As seen from  the phase diagrams 
in
Figs.~\ref{fig5}(a, b), 
using these fits in Eq.~(\ref{1}) gives  quite reasonable estimates for  the maximal mean roll size
of statistically stable turbulent states, see upper blue
lines in  Figs.~\ref{fig5}(a, b).

We now come to the lower bound of the window of allowed $\Gamma_r$. First note that $\Gamma_r$ cannot be infinitesimally 
small, because, in order to form the rolls, the growth rate of the elliptical instability should be much larger than the viscous damping rate, i.e., $\sigma\gg\nu/H^2$.  
One can obtain a more accurate estimate when considering 
a rectangular region that frames a particular roll.
Under the  assumption that the velocity components achieve their maximum at the boundaries of this rectangle or vanish there, 
%there exists a certain constant $c>0$ such that $c^2\overline{{\bf u}^2}\le \Gamma_r^2H^2\overline{(\nabla{\bf u})^2}$. This generalization of the Friedrich inequality  (see the Supplementary Information for a derivation) 
there exists a certain constant $c>2$ such that
$c^2\overline{{\bf u}^2}\le \Gamma_r^2H^2\overline{(\nabla{\bf u})^2}$. 
We call this relation the generalized Friedrichs inequality and derive it in the supplementary material. 
This inequality 
gives as estimate for the lower bound of $\Gamma_r$ 
 \begin{equation}
\Gamma_r\geq c\sqrt{B},
 \label{2}
 \end{equation}
which is plotted as lower blue lines in Fig.~\ref{fig5}(a, b) for $c=9$. 
As can be seen, 
the theoretical slopes reflect the general tendency of the numerical results. Note however  
that at this point we cannot calculate the absolute value of the transition, i.e., the value of $c$. 
We also remark 
that in the case when the container is too slender to keep 2 rolls (with the size, according to (\ref{2})), i.e., when $\Gamma<2c\sqrt{B}$, the flow can take the form of a zonal flow only. Thus, in the no-slip case, the zonal flow configuration \cite{van2014effect,goluskin2014} is possible, but only in small-aspect ratio  containers.

Our approach  also leads to reasonable estimates for the 
window of allowed statistically stable states 
 in the {\it free-slip case}, which was numerically analyzed in ref.~\cite{wang2020zonal}: 
 Taking the dependences $Nu(Ra,Pr)$ and $Re(Ra,Pr)$ for the various roll states from that work, 
 with the definition (\ref{B}) 
 we obtain $B(Ra,Pr)$, 
see Figs.~\ref{fig4}(c, d).  The value of $B$ is always  larger than 0.03. 
Therefore (as discussed above), in the free-slip case,
 the relation (\ref{sigma}) is always fulfilled without any restriction on the maximal roll size.
Thus, very 
 large-$\Gamma_r$ states are possible, while particular state realizations depend on the initial conditions. 
Indeed, this is consistent with
the
numerical  findings in ref.\  
 \cite{wang2020zonal}, which are shown in  Figs.~\ref{fig5}(c, d). Note in particular that in ref.\
 \cite{wang2020zonal} 
% we found  
an as large stable state as $\Gamma_r=64$  (for $\Gamma=128$,  $Ra=10^8$, $Pr=10$)
was found.
 
 What about the lower bound for allowed $\Gamma_r$ in the free-slip case? 
 In contrast to the upper bound, it does exist, and 
%Although, as just seen,  for the free-slip case   the rolls can be very stretched in the horizontal direction, 
%they cannot be of a very small $\Gamma_r$, 
just as in the no-slip case, by arguments
again based on the generalized Friedrich inequality
(see supplementary material),  we can find it. % the  lower bound for $\Gamma_r$. 
In Fig.~\ref{fig5}(c, d)  the result for the smallest $\Gamma_r$
    is plotted.  It is based on the estimate~(\ref{2}) and 
uses the fits from Fig.~\ref{fig4}(c, d) for the smallest values of $B$ with  $c=7$.
Again, the theoretical slopes in the $\Gamma_r - Ra$ and $\Gamma_r - Pr$ phase diagrams 
reflect the general trend  of the
numerical results.

In conclusion, we 
have 
numerically shown the coexistence of multiple statistically stable states
in turbulent RB convection with no-slip BCs, 
with different mean aspect ratios of their  turbulent rolls and different transport properties, even scaling-wise.
We then theoretically illuminated what principles determine the allowed window of
the  mean size of the turbulent convection 
rolls 
(and thus their absolute number),
namely, the existence of  
the
elliptical instability and viscous damping. These criteria also work for the free-slip case. 

Even though a  2D model may  seem somehow artificial, there are various cases in which  
the flow dynamics is mostly 2D, 
 e.g.\ because of geometrical confinement, stratification
or background rotation.  
Therefore our model in itself is relevant, but 
the main ideas of our approach can also be generalized to other wall-bounded turbulent flows, such as 
rotating Rayleigh-B\'enard flow, Taylor-Couette flow, Couette flow with span-wise rotation, 
double diffusive convection, 
etc., and  also to 
geophysical flows such as those mentioned in the introduction. They may also give guidance for  
turbulence flow control, in order
to predict which turbulent states are feasible to be realizable. 

%{\it Acknowledgements:}  
%K.L.~Chong, L.~Liu,  R.~Stevens, and A.\ Tilgner are gratefully acknowledged 
%for discussions and support. 
%We  also acknowledge the Twente Max-Planck Center,
%the Deutsche Forschungsgemeinschaft (Priority Programme SPP 1881 ``Turbulent Superstructures''),
%PRACE for awarding us access to MareNostrum based in Spain at the Barcelona Supercomputing Centre (BSC) under PRACE project number 2017174146. This work was partly carried out on the national e-infrastructure of SURFsara, a subsidiary of SURF cooperation, the collaborative ICT organization for Dutch education and research. Q.W. acknowledges financial support from CSC and NSFC(11621202).
%
    
 {\it Acknowledgements:}  
We  acknowledge the Deutsche Forschungsgemeinschaft (Priority Programme SPP 1881 ``Turbulent Superstructures''). This work was partly carried out on the national e-infrastructure of SURFsara.

%\bibliographystyle{/Users/lohsed/Documents/papers/sty-files/prsty_withtitle}
%\bibliography{/Users/lohsed/Documents/papers/bib-files/literatur}

\bibliographystyle{prsty_withtitle}
\bibliography{literatur}

\begin{thebibliography}{10}

\bibitem{dra81}
P. Drazin and W.~H. Reid, {\em Hydrodynamic stability} (Cambridge University
  Press, Cambridge, 1981).

\bibitem{xi08}
H.-D. Xi and K.-Q. Xia, {\em Flow mode transitions in turbulent thermal
  convection}, Phys. Fluids {\bf 20},  055104  (2008).

\bibitem{poe11}
E.~P. van~der Poel, R.~J. A.~M. Stevens, and D. Lohse, {\em Connecting flow
  structures and heat flux in turbulent {{Rayleigh-B\'enard}} convection},
  Phys. Rev. E {\bf 84},  045303(R)  (2011).

\bibitem{poe12}
E.~P. van~der Poel, R.~J. A.~M. Stevens, and D. Lohse, {\em Flow states in
  two-dimensional {{Rayleigh-B\'enard}} convection as a function of
  aspect-ratio and {{Rayleigh}} number}, Phys. Fluids {\bf 24},  085104
  (2012).

\bibitem{weiss2013}
S. Weiss and G. Ahlers, {\em Effect of tilting on turbulent convection:
  cylindrical samples with aspect ratio $\Gamma= 0. 50$}, J. Fluid Mech. {\bf
  715},  314  (2013).

\bibitem{wang2018}
Q. Wang, Z.-H. Wan, R. Yan, and D.-J. Sun, {\em Multiple states and heat
  transfer in two-dimensional tilted convection with large aspect ratios},
  Phys. Rev. Fluids {\bf 3},  113503  (2018).

\bibitem{xie2018}
Y.-C. Xie, G.-Y. Ding, and K.-Q. Xia, {\em Flow topology transition via global
  bifurcation in thermally driven turbulence}, Phys. Rev. Lett. {\bf 120},
  214501  (2018).

\bibitem{favier2019}
B. Favier, C. Guervilly, and E. Knobloch, {\em Subcritical turbulent condensate
  in rapidly rotating Rayleigh--B{\'e}nard convection}, J. Fluid Mech. {\bf
  864},  R1  (2019).

\bibitem{hui14}
S.~G. Huisman, R.~C.~A. van~der Veen, C. Sun, and D. Lohse, {\em Multiple
  states in highly turbulent {{Taylor-Couette}} flow}, Nat. Commun. {\bf 5},
  3820  (2014).

\bibitem{veen2016}
R.~C. van~der Veen, S.~G. Huisman, O.-Y. Dung, H.~L. Tang, C. Sun, D. Lohse,
  {\it et~al.}, {\em Exploring the phase space of multiple states in highly
  turbulent Taylor-Couette flow}, Phys. Rev. Fluids {\bf 1},  024401  (2016).

\bibitem{ost16}
R. Ostilla-M{\'o}nico, D. Lohse, and R. Verzicco, {\em Effect of roll number on
  the statistics of turbulent Taylor-Couette flow}, Phys. Rev. Fluids {\bf 1},
  054402  (2016).

\bibitem{rav04}
F. Ravelet, L. Mari\'e, A. Chiffaudel, and F. Daviaud, {\em Multistability and
  memory effect in a highly turbulent flow: Experimental evidence for a global
  bifurcation}, Phys. Rev. Lett. {\bf 93},  164501  (2004).

\bibitem{rav08}
F. Ravelet, M. Berhanu, R. Monchaux, S. Aumaitre, A. Chiffaudel, F. Daviaud, B.
  Dubrulle, M. Bourgoin, P. Odier, N. Plihon, J.~F. Pinton, R. Volk, S. Fauve,
  N. Mordant, and F. Petrelis, {\em Chaotic dynamos generated by a turbulent
  flow of liquid sodium}, Phys. Rev. Lett. {\bf 101},  074502  (2008).

\bibitem{cor10}
P. Cortet, A. Chiffaudel, F. Daviaud, and B. Dubrulle, {\em Experimental
  evidence of a phase transition in a closed turbulent flow}, Phys. Rev. Lett.
  {\bf 105},  214501  (2010).

\bibitem{faranda2017}
D. Faranda, Y. Sato, B. Saint-Michel, C. Wiertel, V. Padilla, B. Dubrulle, and
  F. Daviaud, {\em Stochastic chaos in a turbulent swirling flow}, Phys. Rev.
  Lett. {\bf 119},  014502  (2017).

\bibitem{zim11}
D.~S. Zimmerman, A.~A. Triana, and D.~P. Lathrop, {\em Bi-stability in
  turbulent, rotating spherical Couette flow}, Phys. Fluids {\bf 23},  065104
  (2011).

\bibitem{xia2018}
Z. Xia, Y. Shi, Q. Cai, M. Wan, and S. Chen, {\em Multiple states in turbulent
  plane Couette flow with spanwise rotation}, J. Fluid Mech. {\bf 837},  477
  (2018).

\bibitem{bouchet2009random}
F. Bouchet and E. Simonnet, {\em Random changes of flow topology in
  two-dimensional and geophysical turbulence}, Phys. Rev. Lett. {\bf 102},
  094504  (2009).

\bibitem{bouchet2012statistical}
F. Bouchet and A. Venaille, {\em Statistical mechanics of two-dimensional and
  geophysical flows}, Physics reports {\bf 515},  227  (2012).

\bibitem{broecker1985}
W.~S. Broecker, D.~M. Peteet, and D. Rind, {\em Does the ocean--atmosphere
  system have more than one stable mode of operation?}, Nature {\bf 315},  21
  (1985).

\bibitem{schmeits2001}
M.~J. Schmeits and H.~A. Dijkstra, {\em Bimodal behavior of the Kuroshio and
  the Gulf Stream}, J. Phys. Oceanogr. {\bf 31},  3435  (2001).

\bibitem{ganopolski2002}
A. Ganopolski and S. Rahmstorf, {\em Abrupt glacial climate changes due to
  stochastic resonance}, Phys. Rev. Lett. {\bf 88},  038501  (2002).

\bibitem{glatzmaiers1995}
G.~A. Glatzmaiers and P.~H. Roberts, {\em A three-dimensional self-consistent
  computer simulation of a geomagnetic field reversal}, Nature {\bf 377},  203
  (1995).

\bibitem{li2002}
J. Li, T. Sato, and A. Kageyama, {\em Repeated and sudden reversals of the
  dipole field generated by a spherical dynamo action}, Science {\bf 295},
  1887  (2002).

\bibitem{olson2010}
P.~L. Olson, R.~S. Coe, P.~E. Driscoll, G.~A. Glatzmaier, and P.~H. Roberts,
  {\em Geodynamo reversal frequency and heterogeneous core--mantle boundary
  heat flow}, Phys. Earth and Planetary Interiors {\bf 180},  66  (2010).

\bibitem{sheyko2016magnetic}
A. Sheyko, C.~C. Finlay, and A. Jackson, {\em Magnetic reversals from planetary
  dynamo waves}, Nature {\bf 539},  551  (2016).

\bibitem{weeks1997}
E.~R. Weeks, Y. Tian, J. Urbach, K. Ide, H.~L. Swinney, and M. Ghil, {\em
  Transitions between blocked and zonal flows in a rotating annulus with
  topography}, Science {\bf 278},  1598  (1997).

\bibitem{bouchet2019rare}
F. Bouchet, J. Rolland, and E. Simonnet, {\em Rare event algorithm links
  transitions in turbulent flows with activated nucleations}, Phys. Rev. Lett.
  {\bf 122},  074502  (2019).

\bibitem{kol41b}
A.~N. Kolmogorov, {\em On degeneration (decay) of isotropic turbulence in
  incompressible viscous liquid}, Dokl. Akad. Nauk SSSR {\bf 31},  538  (1941).

\bibitem{ahl09}
G. Ahlers, S. Grossmann, and D. Lohse, {\em Heat transfer and large scale
  dynamics in turbulent {{{{Rayleigh-B\'enard}}}} convection}, Rev. Mod. Phys.
  {\bf 81},  503  (2009).

\bibitem{loh10}
D. Lohse and K.-Q. Xia, {\em Small-scale properties of turbulent
  {{Rayleigh-B\'enard}} convection}, Annu. Rev. Fluid Mech. {\bf 42},  335
  (2010).

\bibitem{chi12}
F. Chilla and J. Schumacher, {\em New perspectives in turbulent
  {{Rayleigh-B\'enard}} convection}, Eur. Phys. J. E {\bf 35},  58  (2012).

\bibitem{poe13}
E.~P. van~der Poel, R.~J. A.~M. Stevens, and D. Lohse, {\em Comparison between
  two- and three-dimensional {{Rayleigh-B\'enard}} convection}, J. Fluid Mech.
  {\bf 736},  177  (2013).

\bibitem{poe15cf}
E.~P. van~der Poel, R. Ostilla-M\'onico, J. Donners, and R. Verzicco, {\em A
  pencil distributed finite difference code for strongly turbulent
  wall--bounded flows}, Computers \& Fluids {\bf 116},  10  (2015).

\bibitem{shi10}
O. Shishkina, R.~J. A.~M. Stevens, S. Grossmann, and D. Lohse, {\em Boundary
  layer structure in turbulent thermal convection and its consequences for the
  required numerical resolution}, New J. Phys. {\bf 12},  075022  (2010).

\bibitem{kooij2018}
G.~L. Kooij, M.~A. Botchev, E.~M. Frederix, B.~J. Geurts, S. Horn, D. Lohse,
  E.~P. van~der Poel, O. Shishkina, R.~J. A.~M. Stevens, and R. Verzicco, {\em
  Comparison of computational codes for direct numerical simulations of
  turbulent Rayleigh--B{\'e}nard convection}, Computers \& Fluids {\bf 166},  1
   (2018).

\bibitem{zhu18b}
X. Zhu, V. Mathai, R.~J. A.~M. Stevens, R. Verzicco, and D. Lohse, {\em
  Transition to the Ultimate Regime in Two-Dimensional {{Rayleigh-B\'enard}}
  Convection}, Phys. Rev. Lett. {\bf 120},  144503  (2018).

\bibitem{zhu2019reply}
X. Zhu, V. Mathai, R.~J. A.~M. Stevens, R. Verzicco, and D. Lohse, {\em {Reply
  to “Absence of evidence for the ultimate regime in two-dimensional
  Rayleigh-B{\'e}nard convection.”}}, Phys. Rev. Lett. {\bf 123},  259402
  (2019).

\bibitem{landman1987}
M. Landman and P. Saffman, {\em The three-dimensional instability of strained
  vortices in a viscous fluid}, Phys. Fluids {\bf 30},  2339  (1987).

\bibitem{waleffe1990}
F. Waleffe, {\em On the three-dimensional instability of strained vortices},
  Phys. Fluids A {\bf 2},  76  (1990).

\bibitem{kerswell2002}
R.~R. Kerswell, {\em Elliptical instability}, Annu. Rev. Fluid Mech. {\bf 34},
  83  (2002).

\bibitem{zwirner2020}
L. Zwirner, A. Tilgner, and O. Shishkina, {\em Elliptical Instability and
  Multi-Roll Flow Modes of the Large-scale Circulation in Confined Turbulent
  Rayleigh-B\'enard Convection}, arXiv preprint arXiv:2002.06951  (2020).

\bibitem{shr90}
B.~I. Shraiman and E.~D. Siggia, {\em Heat transport in high-{{Rayleigh}}
  number convection}, Phys. Rev. A {\bf 42},  3650  (1990).

\bibitem{van2014effect}
E.~P. van~der Poel, R. Ostilla-M{\'o}nico, R. Verzicco, and D. Lohse, {\em
  {Effect of velocity boundary conditions on the heat transfer and flow
  topology in two-dimensional Rayleigh-B{\'e}nard convection}}, Phys. Rev. E  .

\bibitem{goluskin2014}
D. Goluskin, H. Johnston, G.~R. Flierl, and E.~A. Spiegel, {\em Convectively
  driven shear and decreased heat flux}, J. Fluid Mech. {\bf 759},  360
  (2014).

\bibitem{wang2020zonal}
Q. Wang, K.-L. Chong, R.~J. A.~M. Stevens, R. Verzicco, and D. Lohse, {\em
  {From zonal flow to convection rolls in Rayleigh-B\'enard convection with
  free-slip plates}}, arXiv:2005.02084  .

\end{thebibliography}

\end{document}